\documentclass[12pt]{article}
\usepackage{doublespace}
\usepackage{citesort}
\usepackage{array}
\begin{document}
\newcommand{\eq}[1]{eq.~\ref{#1}} 
\newcommand{\Eq}[1]{Eq.~\ref{#1}} 
\newcommand{\smq}{\mbox{$\simeq$}} 
\newcommand{\npi}{\hspace{-0.5cm}} 
\newcommand{\apl}[1]{Appl. Phys. Lett. {\bf {#1}}}
\newcommand{\prb}[1]{Phys. Rev. B {\bf {#1}}}
\newcommand{\prl}[1]{Phys. Rev. Lett. {\bf {#1}}}
\begin{spacing}{1.3}

\begin{center}
{\large
{\bf 
Optical spectroscopy of single quantum dots at tunable positive, 
neutral and negative charge states
}
}

\vspace{1cm}

D. V. Regelman, E. Dekel, D. Gershoni and E. Ehrenfreund\\
Physics Department and Solid State Institute, 
Technion--Israel Institute of Technology, Haifa 32000, Israel 

A.J.Williamson, J. Shumway, A. Zunger\\
National Renewable Energy Laboratory, Golden CO 80401, USA

W.V. Schoenfeld and P.M. Petroff\\
Materials Department, University of California,
Santa Barbara, CA 93106, USA
\end{center}

\vspace{1cm}

\today

\vspace{1cm}

We report on the observation of photoluminescence from 
positive, neutral and negative charge states of single semiconductor quantum dots.
For this purpose we designed a structure enabling optical injection of a
controlled unequal number of negative electrons and positive holes into an
isolated InGaAs quantum dot embedded in a GaAs matrix. Thereby, we optically
produced the charge states -3, -2, -1, 0, +1 and +2. The injected carriers
form confined collective 'artificial atoms and molecules' states in the quantum dot.
We resolve spectrally and temporally the photoluminescence from an optically
excited quantum dot and use it to identify collective states, which contain
charge of one type, coupled to few charges of the other type. These states
can be viewed as the artificial analog of charged atoms such as H$^{-}$, 
H$^{-2}$, H$^{-3}$, and charged molecules such as H$_{2}^{+}$ and 
H$_{3}^{+2}$. Unlike higher dimensionality systems, where negative or 
positive charging always results in reduction of the emission energy due 
to electron-hole pair recombination, in our dots, negative charging reduces 
the emission energy, relative to the charge-neutral case, while 
positive charging increases it.
Pseudopotential model calculations reveal that the enhanced spatial
localization of the hole-wavefunction, relative to that of the electron 
in these dots, is the reason for this effect.

\vspace{1cm}


\newpage

\vspace{5mm}

\section
{\large \npi 
{{\bf I. Introduction}}
}
\vspace{2mm}


Real atoms and molecules often appear in nature as charged ions. The
magnitude and sign of their ionic charge reflects their propensity to
successively fill orbitals (the Aufbau principle) and to maximize spin
(Hund's rule). As a result they often exhibit only unipolarity (being either
negative or positive) and remain in a very restricted range of charge values
(e.g. $O^{0},O^{-1},O^{-2}$ or $Fe^{+2},Fe^{+3}$, etc.) even under extreme
changes in their chemical environments. 

Semiconductor quantum dots (QDs) are of fundamental and technological
contemporary interest mainly for their similarities to the fundamental
building blocks of nature, (they are often referred to as ''artificial
atoms'' \cite{1,2,3}) and because they are considered as the basis for new
generations of lasers \cite{4}, memory devices \cite{5}, electronics \cite{6}
and quantum computing \cite{7}. 

Unlike their atomic analogs, however, semiconductor QDs exhibit large
electrostatic capacitance, which enables a remarkably wide range of charge
states. This has been demonstrated by carrier injection either
electronically \cite{6,8,9} by scanning tunneling microscopy \cite{10}, or
even optically \cite{11,12}. This ability to variably charge semiconductor
QDs makes it possible to use them as a natural laboratory for studies of inter-electronic
interactions in confined spaces and at the same time it creates the basis
for their potential applications. Two physical quantities are of particular
importance when quantum dot charging is considered: (i) The charging energy, 
$\mu _{N}=E_{N}-E_{N-1}$ required to add a carrier to a QD containing
already $N-1$ ''spectator'' carriers. Measurements \cite{2,6,8,10} and
calculations \cite{13,14} of QD charging energies have revealed deviations
from the Aufbau principle and Hund's rule, in contrast with the situation in
real atoms \cite{14}. (ii) The electron - hole (e-h) recombination energy 
$\Delta E_{eh}(N_{e},N_{h})$ in the presence of $N_{e}-1$ and $N_{h}-1$
additional ``spectator'' electrons and holes. ($\Delta E_{eh}$ is of great importance
for research and applications which involve light, because light emission
from QDs originates from the recombination of an e-h pair). Measurements of
$ \Delta E_{eh}(N_{e},N_{h})$ were so far carried out only for neutral \cite
{15} and negatively charged \cite{8,11,16} QDs. It was found that upon
increasing the number of electrons and e-h pairs, $\Delta E_{eh}(N_{e},N_{h})$ 
decreases, i.e. the emission due to the e-h radiative recombination shifts
to the red. 

In higher dimensional systems adding negative and/or positive charge
produces the same effect. This is the case for the positively charged
molecule $H_{2}^{+}$ vs. $H_{2}$ \cite{17} in three dimensions (3D), and the
positively or negatively charged exciton (the $X^+$ or $X^-$ trion) vs. the
exciton, in 2D quantum wells (QWs) \cite{18}. 

In the present study, we designed a device enabling optical injection of a
controlled unequal number of electrons and holes into an isolated InGaAs QD,
producing thus the charge states -3, -2, -1, 0, +1 and +2. The injected
carriers form confined collective 'artificial atoms and molecules' states in
the QD. Radiative e-h pair recombination takes place after such a collective
few carrier state relaxes to its ground state. We resolve spectrally and
temporally the photoluminescence (PL) from the QD and use it to determine
the collective carriers' state from which it originated. In particular, we
identify collective states, which contain charge of one type coupled to few
charges of the other type. These states can be viewed as the artificial
analog of charged atoms such as $H^{-},H^{-2},H^{-3}$, and charged molecules
such as $H_{2}^{+}$ and $H_{3}^{+2}$ \cite{17}. We demonstrate that, unlike
the case in negatively charged QDs, and unlike higher dimensionality charged
collective states, positively charged dots show an increase in $E_{eh}$ so
the emission due to radiative recombination of an e-h pair shifts to the
blue. 

Many-body pseudopotential calculations reveal that this trend originates
from the larger spatial localization of the hole-wavefunction with respect
to that of the electron. The calculations tend to agree with the measured
results. They can essentially be summarized as follows: The energy
associated with hole-hole repulsion is larger than the energy associated
with electron-hole attraction, while the latter is larger than the energy
associated with electron-electron repulsion \cite{19}. Therefore, in the
presently studied QDs, as already alluded to by Landin et al \cite{13},
excess electrons decrease (red shift) the recombination energy, while excess
holes increase (blue shift) it.

\vspace{5mm}

{\large \npi 
{{\bf II. Samples and experimental results}}
}
\vspace{2mm}

We have prepared the semiconductor QDs by self-assembly. We used molecular
beam epitaxy (MBE) to grow InAs islands on GaAs by exploiting the 7\%
lattice mismatch strain driven change from epitaxial to island-like growth
mode \cite{20}. This growth mode change occurs after the GaAs surface is
covered with 1.2 monolayers of InAs. During the growth of the InAs islands,
the sample was not rotated. Therefore, the density of the islands varies
across the sample surface, depending on the distance from the In and As MBE
elemental sources. In particular, one can easily find low-density areas, in
which the average distance between neighboring islands is $\smq \mu m$. 
This distance, which is larger than the resolution limit set by the
diffraction at the optical emission wavelength of these islands, permits PL
spectroscopy of individual single islands \cite{15}.

The self-assembled InAs islands were partially covered by deposition of a
2-3 nm thick epitaxial layer of GaAs. The growth was then interrupted for a
minute, to allow melting of the uncovered part of the InAs islands and
diffusion of In (Ga) atoms from (into) the strained islands \cite{21}. A
deposition of GaAs cap layer terminated the growth sequence. The QDs thus
produced are of very high crystalline quality, their typical dimensions are
~40 nm base and ~3 nm height. They form deep local potential traps for both
negative (electrons) and positive (holes) charge carriers. 

Our self-assembled quantum dots (SAQDs) \cite{15,22} and similar other QD
systems were very intensively studied recently using optical excitation and
spectral analysis of the resulting PL emission from single QDs 
\cite{3,8,11,13,16,23,24}. These studies established both experimentally and
theoretically that the number of carriers, which occupy a photoexcited QD,
greatly determines its PL spectrum. 

Since optical excitation is intrinsically neutral (electrons and holes are
always photogenerated in pairs), it is not straightforward to use it for
investigating charged QDs \cite{8}. Two innovative methods have been
recently used to optically charge QDs. The first utilizes spatial separation
of photogenerated electron hole (e-h) pairs in coupled narrow and wide GaAs
QWs, separated by a thin AlAs barrier layer \cite{5}. In this case, the
lowest energy conduction band in the AlAs barrier (X band) is lower than the
conduction band of the narrow QW, but higher than that of the wide one. As a
result, electrons preferentially tunnel through the barrier and accumulate
in the SAQDs within the wider well, while holes remain in the narrow QW (see
Fig. 1a). The second method utilizes capture of photogenerated electrons by
ionized donors (see Fig. 1b-d) \cite{11}. In this case, the holes quickly
arrive at the QDs, while the captured electrons slowly tunnel or hope from
the donors to the QDs, thereby effectively charging or discharging the QDs
during the photoexcitation. In this work, we combined these two methods to
effectively use the excitation density in order to control the number of electrons
present in the QD when radiative recombination occurs (see Fig. 1). 

Two samples were investigated. Sample A, which is used here as a control,
neutral sample, consists of a layer of low density In(Ga)As SAQDs embedded
only within a thick layer of GaAs \cite{22}. Sample B, which we used for
optical charging, consists of a layer of similar SAQDs, embedded within the
wider of two coupled GaAs QWs, separated by a thin AlAs barrier 
layer \cite{5}. 

We spatially, spectrally and temporally resolved the PL emission from single
SAQDs in both samples using a variable temperature confocal microscope
setup \cite{15}. 
Fig. 2 compares the pulse excited PL spectra of
the neutral (Fig. 2a) and mixed type (Fig. 2b) samples. 
The excitation in both pulse and cw (not shown here) excitations was at photon energy of 1.75 eV
and the repetition rate of the picosecond pulses was 78 MHz (\smq13 ns
separation between pulses)
In Fig. 3 we present the spectrally integrated PL emission intensity from 
the various spectral lines of Fig. 2, as a function of the excitation 
intensity for cw (Fig. 3a and 3c) and pulse excitation (Fig. 3b and 3d).
By comparing the PL spectra from the neutral sample (A) to
that from the charged one (B) and by comparing the cw PL spectra to its
pulsed evolution as we increased the excitation
power, we identified the various discrete spectral lines in the spectra.
They are marked in Fig. 2 by the collective carriers' state from which they
resulted. 

\vspace{5mm}

{\large \npi 
{{\bf III. Discussion}}
}
\vspace{2mm}

We interpret the evolution with excitation intensity as follows: At very low
cw excitation intensities of neutral QD, there is a finite probability to
find only one electron-hole (e-h) pair at their lowest energy 
levels, $ e_{1},h_{1}$ (the analog of an S shell of atoms) within the QD. The
radiative recombination of the e-h pair (exciton) gives rise to the PL
spectral line, which we denote by $X^{0}$ (we ignore in this discussion the
"fine" exciton structure \cite{25a}, which gives rise to the fraction of a meV spectral
line splittings observed in Fig. 2). With the increase in the
excitation power, the probability to find few e-h pairs within the QD
increases significantly. Since each level in the QD can accommodate only two
carriers, when three e-h pairs are confined in the QD, the carriers must
occupy the first ($e_{1},h_{1}$) and second ($e_{2},h_{2}$) single-particle
energy levels. Recombination of carriers from the second level gives rise to
the group of spectral lines, which we denote by P, in analogy with the P
shell of atoms. The expected four-fold (including spin) degeneracy of the 
P level of our QDs is removed by the
QDs asymmetrical shape. Therefore the spectral range which we denote by 
P contains two subgroups of lines,
typically\cite{22} roughly six meV apart. The magnitude of this splitting is in agreement with our theoretical 
model (see below).
At these excitation intensities, satellite spectral lines appear on the
lower energy side of the $X^{0}$ line. These lines are due to the
exchange energies between pairs of same charge carriers that belong to
different single particle energy levels (shells). The exchange interaction,
reduces the bandgap of the QD (in a similar way to the well known bulk
phenomenon of "bandgap renormalization"), and, with the increase in the
excitation power, it gives rise to subsequently red shifted 
satellite PL lines ($nX^0$), 
as higher numbers of spectator e-h pairs are present during the radiative 
recombination \cite {15,22,23}. 

Immediately after the optical pulse, excitation intensity
dependent number of photogenerated e-h pairs ($N_{x}$) reaches the QD. Their
radiative recombination process is sequential, and the e-h pairs recombine
one by one. Therefore, all the pair numbers that are smaller than $N_{x}$
contribute to the temporally integrated PL spectrum. This typical behavior
is demonstrated in Fig. 2a in which the temporally integrated PL spectra from
a single QD of the control sample (A) are presented for
various excitation powers. As can be seen in the figure, the 
PL intensity of all the spectral lines
reach maximum and remains constant for further increase in the excitation power.
This behavior is well
described by a set of coupled rate equations model \cite{22}
which can be analytically solved to yield the probabilities to find 
the photoexcited QD occupied by a given number of e-h pairs \cite {22}. 
In Fig. 3b the measured PL intensity of the $X^{0}$ line as a function of the 
pulse-excitation power is compared with the calculated \cite {22} 
number of $X^0$ emitted photons as a function of the 
number of photogenerated excitons in the QD, for each pulse.
We note here that a necessary condition for the above analysis to
hold is that the pulse repetition rate is slow enough, such that all the
photogenerated pairs recombine before the next excitation pulse arrives. 

When cw excitation is used, the evolution of the temporally integrated PL
spectra with the increase in excitation intensity is different. 
In this situation, 
the probability to find a certain number of e-h pairs within the QD
reaches a steady state.
The higher the excitation power is the higher is the probability to find
large numbers of e-h pairs in the QD, while the probability to find the 
QD with a small
number of pairs rapidly decreases. 
As a result, all the observed discrete PL lines at their appearance order undergo a cycle in
which their PL intensity first increases, then it reaches a well defined maximum, 
and eventually it significantly weakens. 
In Fig. 3a we compare the measured spectrally integrated PL intensity of 
the $X^{0}$ line as a function of the excitation power, 
with its calculated emission rate as a function of the photogeneration 
rate of excitons in the QD. 

The PL spectrum from the mixed-type sample B, and its evolution with 
excitation power is very different. The reason for this difference is
the fact that the SAQDs in the mixed-type sample B are initially charged
with electrons \cite{25}. These electrons are accumulated in the SAQDs due to the
sample design, which facilitates efficient hopping transport of electrons
from residual donors. The maximal number of electrons in a given QD
is limited by the electrostatic repulsion, which eventually forces
new electrons to be unbound, i.e. have energies above the wetting layer
continuum onset. We found experimentally
that the maximal number of electrons is three, 
in excellent agreement with our model calculations (see below).

With the optical excitation, the ionized donors separate a certain fraction
of the photogenerated e-h pairs. Then, while the donors capture electrons
and delay their diffusion \cite{11}, the holes quickly reach the QDs and
deplete the QDs initial electronic charge. The recombination of an e-h pair
in the presence of a decreasing number of electrons reveals itself in a
series of small discrete lines to the lower energy side of the PL line $X^{0}$
(Fig. 2b). At very low excitation intensity, a single sharp PL line
(which we denote in Fig.2b by $X^{-3})$ appears. The $X^{-3}$ line originates
from the recombination of an e-h pair in the presence of additional 3 spectator electrons.
As can be seen in Fig. 2b, with the increase in excitation power, new spectral lines
appear on the higher energy side of the first to appear line.
With increasing intensity, the $X^{-3}$ line weakens and there emerge small higher
energy lines (the lowest energy of which is marked by $X^{-2}$) 
which originates from pair recombination in the
presence of two additional electrons. With further increase of the excitation power
these lines lose strength as well and few other spectral lines,
at yet higher energy, appear. We mark the strongest line in this group by $X^{-1}$.
At yet higher excitation intensity, a spectral line, which we denote by $X^{0}$ emerges.
With further increase of the excitation power, new satellite PL lines appear, but now they 
appear in both the higher and lower energy sides of the neutral, $X^{0}$ line of
sample B. While the lines to the lower energy side are similar to those observed in the PL 
spectra
from QDs in the control sample (the $nX^0$ lines), the higher energy lines are entirely 
different. 
This difference results from the mechanism of preferential hopping of photoexcited holes
into the QDs. This mechanism leads, at high enough intensities, not only to
negative charge depletion, but also to positive charging of the QDs.

We thus identify the higher energy satellites of the $X^0$ line 
as resulting from e-h recombination in the presence of additional holes within the QD.
We attribute the first two higher energy satellites 
to final states with one (line $X^{+1}$) and two (line $X^{+2}$) holes. 
The positive charging mechanism at high optical excitation power should not come as a surprise,
since while the SAQD cannot contain more than 3 electrons, it can definitely collect holes 
from a larger number of neighboring photodeionized donors.
The high energy side of the $X^{0}$ line does not extend over 6 meV 
and we never observed more than two discernible satellites at this side of the spectrum.
As explained below, this does not mean that the QD cannot be charged with 
a larger number of holes. It reflects the fact that the observed blue shift is maximal
for a positively charged QD with only two holes. 

The evolution of the spectra with the increase 
in pulse-excitation power (Fig. 2b) is similar to that of the control sample
except for one important difference. Here, the PL intensity of 
all the spectral lines due to recombination from negatively charge states, which 
appear at low excitation power prior to the appearance of the $X^0$ line, 
reaches maximum and then considerably weakens at higher
excitation power. The $X^0$ line is the first spectral line
that behaves differently. Its intensity reaches a maximum and remains
constant as the excitation power is further increased.

In Fig. 3c (3d) we present the spectrally integrated PL intensity of various 
spectral lines from a single mixed type QD as a function of the cw (pulse) 
excitation intensity. The figure demonstrates that the
``neutral'' PL lines, such as $X^0$ and $nX^0$, evolve 
similarly to those of sample A. Under pulse excitation their intensity reaches 
a maximum and remains unchanged  as the pulse excitation intensity is 
further increased. At the same time, the negatively ``charged''
PL lines evolve like the lines of sample A under cw-mode excitation. 
We attribute this difference to the long hopping times of the trapped photoexcited
electrons, which determine the lifetime of the emission from the various 
charged states. This lifetime can be crudely estimated from the intensity ratio 
between the maximum intensity of the $X^{0}$ PL line under cw excitation and 
that from the negatively charge states (~10:1, see Fig 4c), since the emission intensity at 
maximum is inversely proportional to the state lifetime \cite {22,25b}. 
From the measured decay time of the $X^{0}$ line (\smq1.3 ns, not shown) we
deduce the hopping times and find them to be slightly longer than the pulse
repetition rate (\smq13 ns). Therefore, under pulsed excitation, once the QD is optically
depleted from its initial charge, it remains so for times longer than the time difference
between sequential pulses. Thus, PL lines that result from exciton recombination
in the presence of negative charge evolve like neutral lines under cw-mode excitation
(Fig. 3a), and their intensity weakens with the increase in excitation power.
We use this behavior to sort out the neutral states PL emission 
from that of negatively charged ones, in general, and for identifying
the $X^0$ line in particular. 

The larger is the number of de-ionized donors which participate
in the depletion process, the shorter is the lifetime of the charge depleted 
state that they generate. This can be straightforwardly deduced from Fig. 3d
and more quantitatively by simple rate equation model simulations \cite{25b}.
Since four and five deionized donors are involved in generating the
$q=+1$ and $q=+2$ charge states, respectively, the lifetime of these
states is shorter than that of the charge-neutral state and shorter
from the pulse repetition time.
Hence, the evolution of the PL intensity of the $X^{+1}$ and $X^{+2}$
lines with increasing pulse excitation power, is similar to that
of the $X^0$ line.

\vspace{5mm}

{\large \npi 
{{\bf IV. Comparison with Theory}}
}
\vspace{2mm}

The effect of spectator charges on the recombination energy of the
fundamental $e_{1}-h_{1}$ excitonic transitions $\Delta
E_{e_{1},h_{1}}^{(N_{e},N_{h})}$ is theoretically given by \cite{14}: 
\begin{equation}
\Delta E_{e_{1},h_{1}}^{(N_{e},N_{h})}=
\begin{array}{ll}
\left[ \varepsilon
_{e_{1}}-\varepsilon _{h_{1}}-J_{e_{1},h_{1}}\right]
 -\left[
\sum_{i=2}^{N_{e}}(J_{e_{1},e_{i}}-J_{h_{1},e_{i}})+
\sum_{j=2}^{N_{h}}(J_{h_{1},h_{j}}-J_{e_{1},h_{j}})\right] \\
 +\left[ \Delta _{exch}^{(N_{e})}+\Delta _{exch}^{(N_{h})}\right] 
 +\Delta _{corr}^{(N_{e},N_{h})} \label{Eq1}\\ 
\end{array}
\end{equation}
In Eq. (1), the first term in square brackets is the recombination energy of
the exciton in the absence of additional carriers. Here $\varepsilon_{e_{1}}
$ ($\varepsilon_{h_{1}}$) is the energy of the first single electron (hole)
level and $J_{e_{1},h_{1}}$ is the e-h pair binding energy. The second
bracketed term (''Coulomb shift'', $\delta E_{coul}$) contains the
difference between the Coulomb repulsion and Coulomb attraction terms
between the recombining exciton and the spectator electrons and holes. In
simple models, in which the electrons and holes have the same single
particle wavefunctions, this term vanishes \cite{15,22,23,24}. However, if
the holes (electrons) are more localized than the electrons (holes), then
the Coulomb term results in red (blue) PL shift upon electron (hole)
charging \cite{19}. The third bracketed term (''Exchange shift'', $\delta
E_{exch}$)) is the change in the exchange energies upon charging. The
exchange term is always negative for both electrons and holes charging 
\cite{22,23,24}. This term is responsible for the multi line PL spectrum, since
open shells in the final state give rise to few spin multiplets whose
energies depend on the spin orientation of the carriers in these shells. The
last term (''correlation shift'' $\delta E_{corr}$)) is due to the
difference in correlation between the interacting many carriers within the
QD. This term is always negative as well \cite{27}. In higher dimensionality
systems, such as quantum wires, wells and bulk semiconductors, the carrier
wavefunctions are delocalized in at least one dimension. Therefore, the
Coulomb shift in Eq. (1) is much smaller than the correlation shift. As a
result, red shifts are anticipated for both negative and positive charging.
In zero dimensional QDs whose dimensions are comparable to and smaller than
the bulk exciton radius, the correlation terms are smaller than the Coulomb
and exchange terms \cite{26}. Thus, in some cases, a positive Coulomb term
may overwhelm the exchange and correlation terms, resulting in blue PL shift
upon QD charging \cite{19}. Clearly, a microscopic calculation is needed to
establish the detailed balance between the various terms of Eq. (1). We used
pseudopotential calculations \cite{19} in order to realistically estimate
the various terms in Eq. (1). The first three terms were obtained via
first-order perturbation theory whereas the last term (the correlation
energy) was obtained via configuration-interaction calculations \cite{14,27}.
The QD calculated here \cite{22}, has a slightly elongated lens shape, with
major and minor axis 45 nm and 38 nm (in the 110 and $\bar{1}$10 directions,
respectively), and a height of 2.8 nm. Both the QD and the two monolayer
wetting layer have a uniform composition of In$_{0.5}$Ga$_{0.5}$As, and they
are embedded in a GaAs matrix. The shape, size, and composition are based on
experimental estimations, and they are somewhat uncertain. 

The pseudopotential treats the alloy atomistically, and it includes spin-orbit
interaction and strain. We include the first six bound electron and hole
states in our configuration-interaction expansion. The calculated S-P shells
splitting 
($\varepsilon _{e_{2}}-\varepsilon _{e_{1}}+\varepsilon
_{h_{2}}-\varepsilon _{h_{1}}$ \smq 37 meV), 
well agrees with the measured
one and so does the energy splitting of the P shell 
($\varepsilon _{e_{3}}-\varepsilon _{e_{2}}+\varepsilon
_{h_{3}}-\varepsilon _{h_{2}}$ \smq 6 meV). 

Fig. 4 shows isosurface plots of the calculated density of probability for
electrons and holes in their three lowest energy states. The electric charge
that these isosurfaces contain amount to 75\%. The figure clearly
demonstrates that the holes are more localized than the electrons.
Quantitatively speaking, the volume of the $e_1$ isosurface in Fig.5
 ($1600 nm^3$) is 3 times larger than that of $h_1$. 
Consequently, the Coulomb shift term in Eq. (1) contributes a red
(blue) shift to the excitonic recombination for negative (positive) charging.

By calculating the carrier addition energies $\mu _{N}$ \cite{14}, we find
the energies 1.460, 1.473, 1.500, and 1.510 eV for electron charging of N
=1, 2, 3, and 4, respectively. Only the first three energies are below the
calculated wetting layer conduction band energy of 1.506 eV. This is in
perfect agreement with the fact that we never observed experimentally higher
then N=3 negatively charged QD. Similar calculations for holes yielded that
the SAQD can hold at least 6 holes. 

In Fig. 5 we compare the measured PL energies due to $e_{1}-h_{1}$ pair
recombination from various charged exciton states with the calculated
emission. The energy is measured relative to the uncharged $e_{1}-h_{1}$
recombination energy. Shaded bars indicate the measured peak positions. The
bar positions are obtained by averaging results of measurements from 6
different QDs, and the width of the bars represents the statistical and
experimental uncertainties. Dashed lines represent the sums of the first 3
terms of Eq. (1), whereas solid lines represent the full calculations (with
the correlation terms). 

As can be seen in Fig. 5 the calculations resemble the optical measurements:
(i) We see a blue shift for hole charging and a red shift for electron charging. 
The computed shifts are underestimated relative to experiment. 
We attribute these discrepancies to uncertainties in SAQD shape, size and composition profile. 
(ii) The blue shift due to hole charging is bound from above. It ceases to increase after two or the most three
positive charges (depending on wether the last one is an Aufbau or non-Aufbau state). 
For additional positive charging, the exchange and correlation terms overhelme the Coulomb
term. 
(iii) We calculate two PL lines for q = -2, arising from
the exchange splitting between the triplet and singlet states of the two electrons. The higher energy
line which results from the transition to the triplet state is about
three times stronger \cite{8}.
In the experiment, due to the fact that lines due to few charge states are observed together, 
only the lowest energy line could be safely indentified.
(iv) For the q = -3 we calculate a multi-line PL spectrum, 
corresponding to the non-Aufbau $(e^{1}_{1}h^{1}_{1})(e_{1}e_{2}e_{3})$ initial state
configuration. For the model QD, this configuration is somewhat lower in energy
than the Aufbau-like $(e^{1}_{1}h^{1}_{1})(e_{1}^{1}e_{2}^{2}e_{3}^{0})$initial state
since the calculated inter-electronic exchange energy exceeds the single particle energy
difference. In the experiment, only a single PL line ($X^{-3}$) is always observed, indicating
that the Aufbau-like state is the lower energy one. 
We attribute this discrepancy to more than a factor of two overestimated exchange energies 
(probably due to the above mentioned uncertainties). The influence of near by local charges
may also contribute to this discrepancy.

Fig. 5 demonstrates a fundamental difference between zero dimensional
charged excitons and those confined in higher dimensional systems. The
negatively charged excitons in our SAQDs, like their free charged atomistic
analog \cite{17}, have lower recombination energies than their corresponding
neutral complexes. Positively charged QD excitons, unlike their analog free
positively charged molecules \cite{17}, have larger recombination energies.
This novel observation may prove to be very useful in future applications
of semiconductor quantum dots, where their optical emission can be
discretely varied by controlled carrier injection. 

{\bf Acknowledgments}: The research was supported by the US-Israel
Science Foundation (453/97) and by the Israel Science Foundation founded by
the Israel Academy of Sciences. 


\bigskip

\textbf{Figure Caption:}
\begin{description}

\item[Fig. 1: ]
Schematic description of the mixed type SAQD sample. (a) Initial
charging of the QD with electrons from ionized donors. (b) e-h pair
photogeneration (c) QD photodepletion and (d) QD slow recharging by the
captured electron from the ionized donor.
\item[Fig. 2: ]
Picosecond pulse excited photoluminescence spectra of single QDs for increasing
excitation intensities. (a) From a charge-neutral QD (the control sample A)
and (b) from a charged QD (mixed type sample B). The discrete PL lines due
to the recombination of neutral excitons ($X^{0}$, $nX^{0}$), negatively
charged ($X^{-i}$) and positively charged ($X^{+i}$) excitons are marked in
the figures. 
\item[Fig. 3: ]
Spectrally integrated photoluminescence intensity of the various spectral lines as a
function of excitation intensity (in the integration, possible "fine" structure of the spectral
lines is ignored and background due to neighboring lines is subtracted using a line a gaussian line fitting procedure). 
(a) The neutral ($X^{0}$) line from the control sample A at cw excitation. (b) Same as (a) at pulsed excitation. 
The solid lines in the figures (upper and right hand side axes, where $\tau _{0}$
is the radiative lifetime of $e_{1}-h_{1}$ pair) represent our rate equation
model calculations \cite{22}. Note the typical difference between cw and
pulsed excitation. Whereas in the first case the PL intensity goes through a
clear maximum and then it decreases with further increase of the excitation
density, in the second case it saturates and remains constant. The intensity
dependence of the negatively charged and the neutral PL lines from the mixed
type QD sample B, for cw and for pulse excitations are given in (c) and (d)
respectively. Note the clear distinction between the intensity dependence of
the PL lines due to negatively charged states and those from neutral states.
We use these differences to determine the origin of the various PL lines (see text). 
\item[Fig. 4: ]
Top view of the calculated single electron and hole wavefunctions
squared for the SAQD under study. The isosurfaces contain 75\%  of the total
charge. The pseudopotential calculations use the linear combination of bulk
bands method \cite{27}. Note that the spatial extent of the electron
wavefunction is larger than that of the hole.
\item[Fig. 5: ]
Comparison between the calculated energy of the $e_{1}-h_{1}$ PL
spectral lines (solid lines) and the experimentally measured energy (shaded
bars) for various negatively and positively charged QD exciton states.
Vertical dashed lines denote calculated peak without correlation. Horizontal
arrows show Coulomb, exchange, and correlation shifts. The measured values
represent statistical average over 6 different dots from the mixed type
sample B. In both experiment and theory the emission energy of negatively
charged excitons is lower in energy than that from neutral excitons, while
that from positively charged excitons is higher. 
\end{description}

\end{spacing}

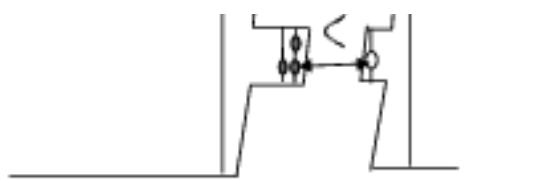
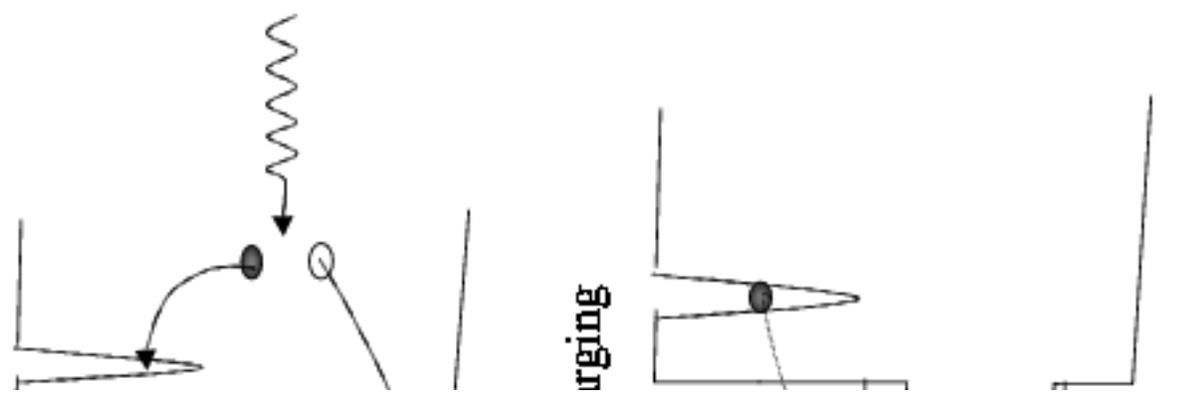
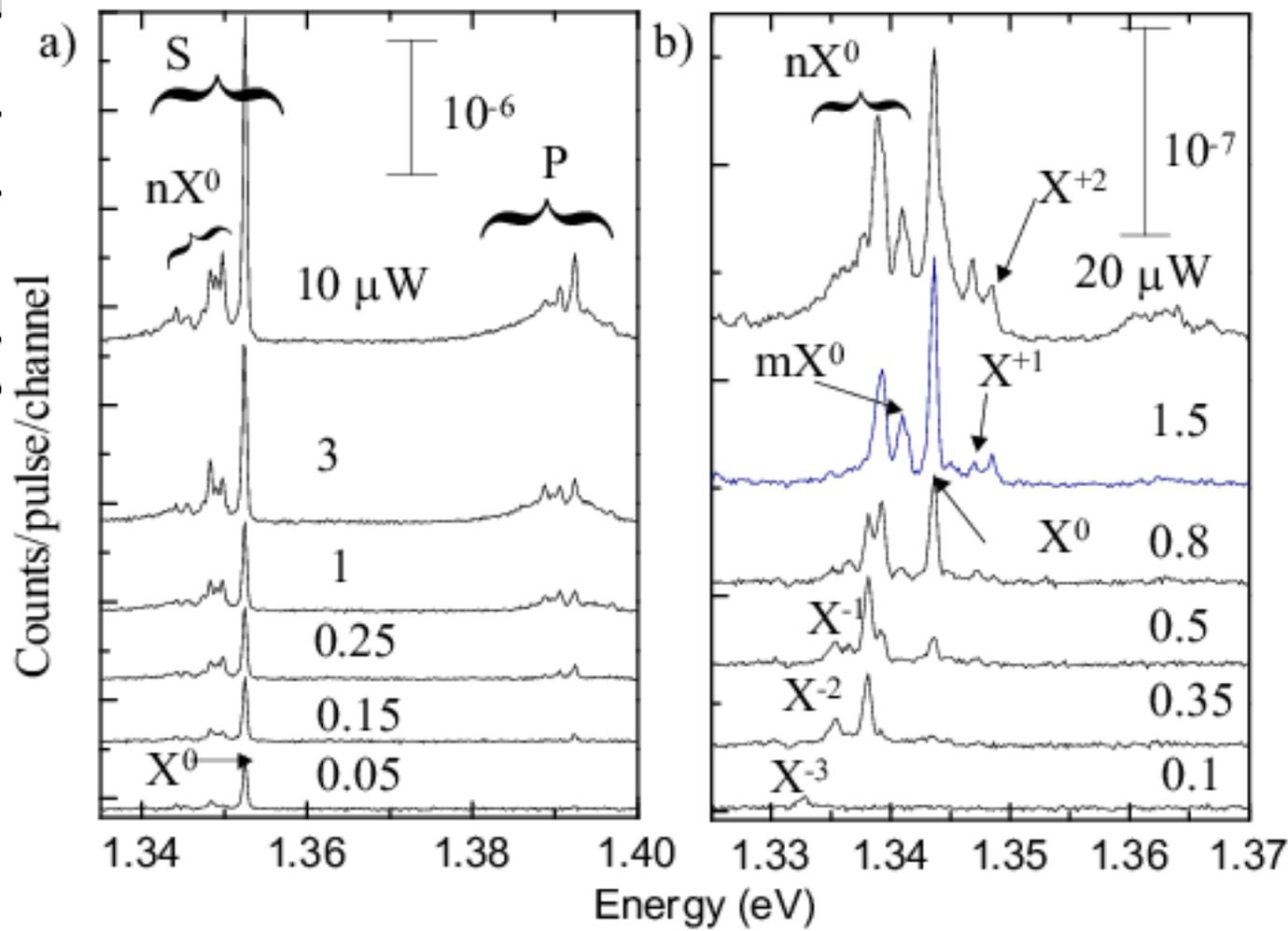
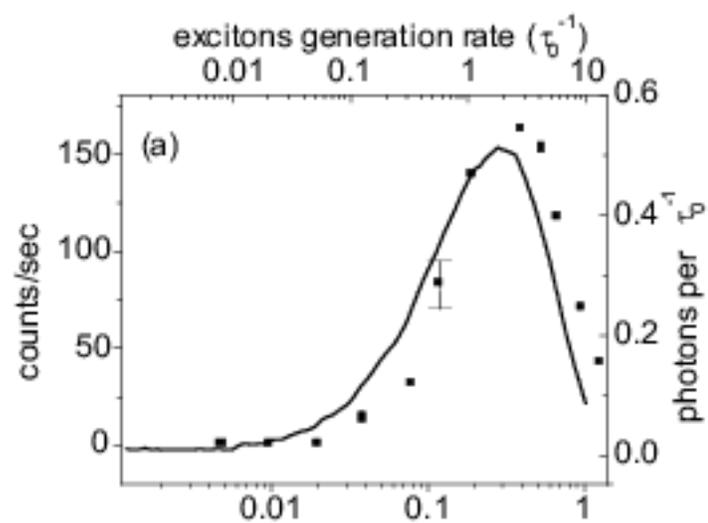
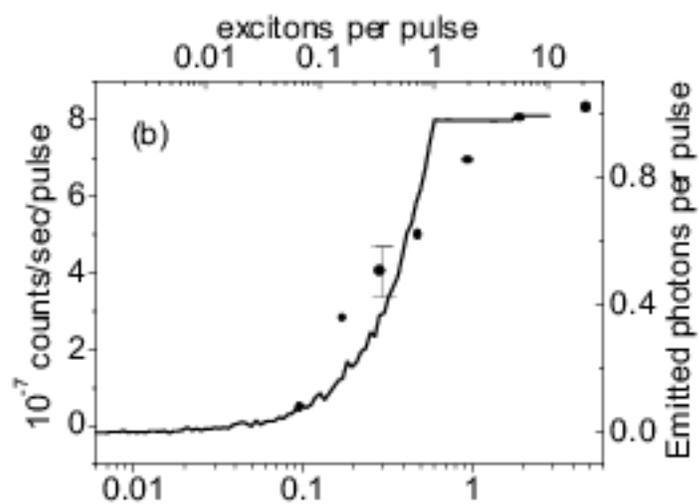
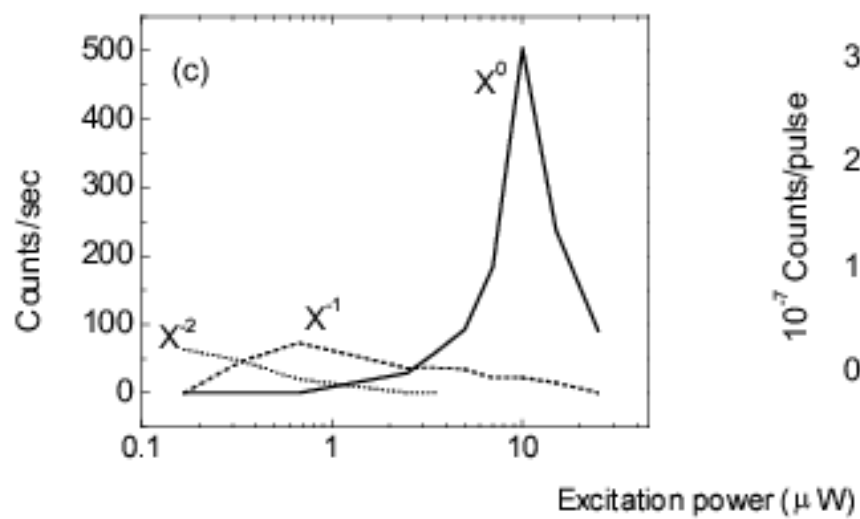
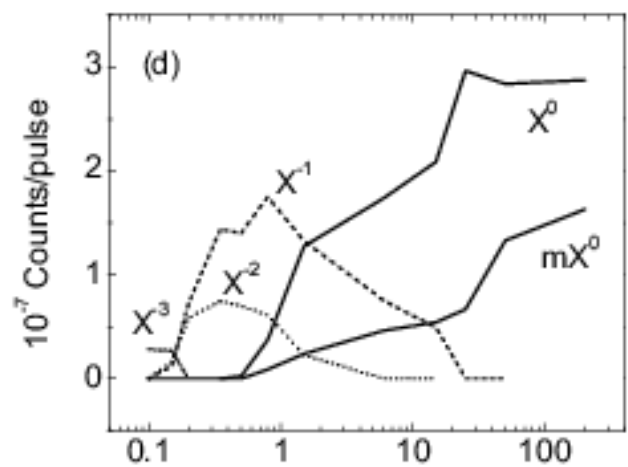
\end{document}